\documentclass[final,3p,times]{elsarticle}
\usepackage{minted}
\usepackage{lineno}

\usepackage{tabularx}
\usepackage{comment}
\usepackage{amssymb}
\usepackage{float}
\usepackage[colorlinks=true,linkcolor=blue,citecolor=blue,urlcolor=blue]{hyperref}
\usepackage{amsmath}
\biboptions{numbers,sort&compress}

\journal{Nuclear Physics A}

\begin{document}

\begin{frontmatter}



\title{Ultracold Neutron Guide-Coating Facility at U.~Winnipeg}



\author[UW]{T.~Hepworth\corref{cor1}\fnref{ILL}} 
\author[UM]{A.~Zahra}
\author[UM]{B.~Algohi}
\author[UW]{R.~de Vries} 
\author[UW]{S.~Pankratz} 
\author[UW]{P.~Switzer} 
\author[UW]{T.~Reimer} 
\author[UW]{M.~McCrea}
\author[UW]{J.W.~Martin}
\author[UW]{R.~Mammei}

\author[TRIUMF]{D.~Anthony}
\author[UNAM]{L.~Barr\'on-Palos}
\author[TRIUMF]{M.~Boss\'e}
\author[USASK]{M.P.~Bradley}
\author[TRIUMF]{A.~Brossard}
\author[UM]{T.~Bui}
\author[TRIUMF]{J.~Chak}
\author[SFU]{R.~Chiba}
\author[TRIUMF]{C.~Davis}
\author[TRIUMF]{K.~Drury} 
\author[TRIUMF]{D.~Fujimoto}
\author[UKYOTO,KURNS]{R.~Fujitani}
\author[UM]{M.~Gericke}
\author[TRIUMF]{P.~Giampa}
\author[TRIUMF]{C.~Gibson}
\author[NCSU]{R.~Golub}
\author[KURNS,RCNP]{T.~Higuchi}
\author[KEK]{G.~Ichikawa}
\author[NAGOYA]{I.~Ide}
\author[RCNP]{S.~Imajo\fnref{RIKEN}}
\author[UM]{A.~Jaison}
\author[UW]{B.~Jamieson}
\author[UW]{M.~Katotoka} 
\author[KEK,SOKENDAI]{S.~Kawasaki}
\author[NAGOYA]{M.~Kitaguchi}
\author[UBC]{W.~Klassen}
\author[UNBC]{E.~Korkmaz}
\author[NCSU]{E.~Korobkina}
\author[UM]{M.~Lavvaf}
\author[TRIUMF]{T.~Lindner}
\author[TRIUMF]{N.~Lo} 
\author[UM]{S.~Longo}
\author[UBC]{K.W.~Madison}
\author[KEK,SOKENDAI]{Y.~Makida}
\author[TRIUMF]{J.~Malcolm} 
\author[UM]{J.~Mammei}
\author[UBCCHEM]{Z.~Mao}
\author[TRIUMF]{C.~Marshall}
\author[TRIUMF,RCNP]{R.~Matsumiya}
\author[UBCCHEM]{E.~Miller}
\author[McGILL]{M.~Miller}
\author[RCNP,KEK,NAGOYA]{K.~Mishima}
\author[UM]{T.~Mohammadi}
\author[UBC,UBCCHEM,TRIUMF]{T.~Momose}
\author[TRIUMF]{M.~Nalbandian} 
\author[KEK,SOKENDAI]{T.~Okamura}
\author[TRIUMF]{R.~Patni} 
\author[TRIUMF,SFU]{R.~Picker}
\author[RCNP,OSAKA]{K.~Qiao}
\author[TRIUMF]{W.D.~Ramsay}
\author[UM]{W.~Rathnakela}
\author[SFU]{D.~Salazar}
\author[NAGOYA]{J.~Sato}
\author[TRIUMF,ORNL]{W.~Schreyer}
\author[RCNP]{T.~Shima}
\author[NAGOYA]{H.M.~Shimizu}
\author[TRIUMF]{S.~Sidhu}
\author[UM]{S.~Stargardter}
\author[TRIUMF]{R.~Stutters} 
\author[RCNP]{I.~Tanihata}
\author[UM]{Tushar}
\author[TRIUMF]{W.T.H.~van~Oers}
\author[TRIUMF]{N.~Yazdandoost}
\author[UBCCHEM]{Q.~Ye}
\author[TRIUMF]{M.~Zhao} 

\cortext[cor1]{Corresponding author. Email: thomashepworth12@gmail.com}
\fntext[ILL]{current affiliation: Physikalisches Institut, Universität Heidelberg, Germany; Institut Laue Langevin, Grenoble, France}
\fntext[RIKEN]{current affiliation: RIKEN Center for Advanced Photonics, Wako, Saitama, Japan}

\affiliation[UW]{organization={Department of Physics, University of Winnipeg},
            addressline={515 Portage Ave}, 
            city={Winnipeg},
            postcode={R3B 2E9}, 
            state={Manitoba},
            country={Canada}}
\affiliation[UM]{organization={Department of Physics and Astronomy, University of Manitoba},
            addressline={424 University Centre}, 
            city={Winnipeg},
            postcode={R3T 2N2}, 
            state={MB},
            country={Canada}}
\affiliation[TRIUMF]{organization={TRIUMF},
            addressline={4004 Wesbrook Mall}, 
            city={Vancouver},
            postcode={V6T 2A3}, 
            state={BC},
            country={Canada}}
\affiliation[UNAM]{organization={Instituto de F\'isica, Universidad Nacional Aut\'onoma de M\'exico},
            addressline={Sendero Bicipuma}, 
            city={Mexico City},
            postcode={04510}, 
            state={CDMX},
            country={M\'exico}}
\affiliation[USASK]{organization={Department of Physics and Engineering Physics, University of Saskatchewan},
            addressline={105 Adminstration Pl}, 
            city={Saskatoon},
            postcode={S7N 5A2}, 
            state={SK},
            country={Canada}}  
\affiliation[SFU]{organization={Department of Physics, Simon Fraser University},
            addressline={8888 University Drive}, 
            city={Burnaby},
            postcode={V5A 1S6}, 
            state={BC},
            country={Canada}}
\affiliation[UKYOTO]{organization={Department of Nuclear Engineering, Kyoto University},
            addressline={Kyotodaigaku-Katsura Nishikyou-ku}, 
            city={Kyoto},
            postcode={615-8540}, 
            country={Japan}}
\affiliation[KURNS]{organization={Institute for Integrated Radiation and Nuclear Science (KURNS), Kyoto University},
            addressline={2, Asashiro-Nishi, Kumatori}, 
            state={Sennan-gun},
            postcode={590-0494}, 
            city={Osaka},
            country={Japan}}

\affiliation[NCSU]{organization={Department of Physics, North Carolina State University},
            addressline={1020 Main Campus Drive}, 
            city={Raleigh},
            postcode={27695}, 
            state={NC},
            country={USA}}
\affiliation[RCNP]{organization={Research Center for Nuclear Physics (RCNP), The University of Osaka},
            addressline={10-1 Mihogaoka}, 
            city={Osaka},
            postcode={567-0047}, 
            state={Ibaraki},
            country={Japan}}
\affiliation[KEK]{organization={High Energy Accelerator Research Organization (KEK)},
            addressline={1-1 Oho}, 
            city={Tsukuba},
            postcode={305-0801}, 
            state={Ibaraki},
            country={Japan}}
\affiliation[NAGOYA]{organization={Department of Physics, Nagoya University},
            addressline={Furocho}, 
            city={Nagoya},
            postcode={464-8601}, 
            state={Aichi},
            country={Japan}}
\affiliation[SOKENDAI]{organization={The Graduate University for Advanced Studies (Sokendai)},
            addressline={
            address}, 
            city={Tsukuba},
            postcode={240-0193}, 
            state={Ibaraki},
            country={Japan}}
\affiliation[UBC]{organization={Department of Physics and Astronomy, University of British Columbia},
            addressline={170-6371 Crescent Road}, 
            city={Vancouver},
            postcode={V6T 1Z2}, 
            state={BC},
            country={Canada}}       
\affiliation[UNBC]{organization={Department of Physics,University of Northern British Columbia},
            addressline={3333 University Way}, 
            city={Prince George},
            postcode={V2N 4Z9}, 
            state={BC},
            country={Canada}}
\affiliation[UBCCHEM]{organization={Department of Chemistry, University of British Columbia},
            addressline={170-6371 Crescent Road}, 
            city={Vancouver},
            postcode={V6T 1Z2}, 
            state={BC},
            country={Canada}}               
\affiliation[McGILL]{organization={Department of Physics, McGill University},
            addressline={845 Sherbrooke Street West}, 
            city={Montreal},
            postcode={H3A 0G4}, 
            state={QC},
            country={Canada}}
\affiliation[OSAKA]{organization={Graduate School of Science, The University of Osaka},
            addressline={1-1 Machikaneyama-cho}, 
            city={Osaka},
            postcode={560-0043}, 
            state={Osaka Prefecture},
            country={Japan}}
\affiliation[ORNL]{organization={Physics Division, Oak Ridge National Laboratory},
            addressline={1 Bethel Valley Road}, 
            city={Oak Ridge},
            postcode={37830}, 
            state={TN},
            country={USA}}

\begin{abstract}
We report the construction and commissioning of a new ultracold neutron (UCN) guide-coating facility at the University of Winnipeg. The facility employs pulsed laser deposition (PLD) to produce diamond-like carbon (DLC) coatings on cylindrical UCN guides up to 1 m in length with a 200 mm outer diameter. DLC is a promising material for UCN transport and storage due to its high real component of the optical potential, low neutron absorption cross section, and low depolarization probabilities. First coating attempts on a full length aluminum UCN guide and matching blank flange were successfully coated with a carbon film with density of 2.3 g/cm$^3$, corresponding to optical potentials of 200 neV, as measured by X-ray reflectometry (XRR). Coating thicknesses were measured to be 90 nm for the UCN guide and 180 nm for the flange with no evidence of delamination. The implementation of a plasma plume collimator and plasma feed back control via a time of flight in vacuum ion probe produced a film with an XRR measured density of 2.8 g/cm$^3$, corresponding to an optical potential of 240 neV. This 80 nm thick film had poor adhesion to the aluminum tube substrate. These results establish a baseline for the coating facility. Ongoing and future work focuses on improving the diamond content of films and adhesion through plasma plume collimation, TOF ion probe feed back, and pre/post treatment methods with the goal of providing high quality DLC UCN guides for the TUCAN experiment at TRIUMF.
\end{abstract}




\end{frontmatter}



\section{Introduction}\label{sec:intro}
Ultracold neutrons (UCNs) are free neutrons with energies up to about 350 neV, corresponding to velocities of up to 8 m/s. UCNs can be stored magnetically or in material bottles via the strong interaction,  making them an excellent probe for precision searches for new physics. This includes measurements of the neutron's permanent electric dipole moment (EDM) \cite{PhysRevLett.124.081803, PanEDM_2019} and lifetime \cite{UCN_tau_2021, Auler_2024}. A new UCN source at TRIUMF has recently produced its first UCNs \cite{TUCAN:2025rxj}, with plans to deliver UCN from the source to experiments 15~m away: the TRIUMF Ultracold Advanced Neutron (TUCAN) EDM experiment and Precision Experiment on Neutron Lifetime Operating with Proton Extraction (PENeLOPE) experiment \cite{Higuchi:2025zlp,materne2009penelope}. \\

UCN containment via strong interactions is characterized by a nuclide-specific pseudo-potential, referred to as the Fermi potential \cite{Gol91, Ign90, Steyerl:2020xpc}. Averaging over many scattering centers one arrives at the optical potential:
\begin{equation}
    V = \frac{2\pi\hbar^2}{m_{\mathrm{n}}}Na,
    \label{eq:V}
\end{equation}
where $m_{\mathrm{n}}$ is the mass of the neutron, $N$ is the number density of nuclei in the material, and $a$ is the coherent bound nuclear scattering length. UCNs are able to reflect at any angle of incidence provided they have a lower energy than the optical potential barrier, corresponding to the critical velocity $v_{\mathrm{c}} =  \sqrt{2V/m_{\mathrm{n}}}$.\\

UCNs have many loss mechanisms when scattering off materials \cite{Gol91, Ign90, Steyerl:2020xpc}. These losses can be accounted for by defining the complex neutron-optical potential $U= V - iW$, with $V$ defined in Eq. (\ref{eq:V}), and the imaginary component
\begin{equation}
    W = \frac{\hbar}{2} N \sigma  v,
    \label{eq:W}
\end{equation}
where $\sigma$ is the total loss cross section per atom of the interacting material, and $v$ is the speed of the neutron. The most ideal materials have high $V$ and low $W$. The neutron loss coefficient $\eta = W/V$ is often used as a metric for characterizing the suitability of materials to be used in UCN experiments \cite{Gol91,Ign90, Steyerl:2020xpc}. In experiments with durations of hundreds of seconds, neutrons can have on the order of 10$^4$ wall interactions. Thus, typical materials which are good for UCN applications have $\eta < 10^{-4}$  \cite{Serebrov_2005, PATTIE201764, PhysRevC.96.035205, korobkina2004temperature, BRYS2005429, brys2006dlc_ucn, heule2007dlc_guides, CYTOP}. The best beryllium coatings have been shown to have $\eta$ = $3\cdot10^{-5}$, but are less used due to being toxic and extremely costly.  Previously, NiP coated UCN guides, which are the baseline for the TUCAN experiment, have been found to have $\eta = (3.5 \pm 0.5)\cdot10^{-4}$ ($V = 214\pm 5$~neV) \cite{AKATSUKA2023168106}. To further maximize the number of UCNs available to experiments, diamond-like carbon (DLC) coatings are being investigated as a replacement for NiP in the storage and transport for experiments at TRIUMF. DLC coatings in the past have been found to have loss coefficients in the $10^{-5} - 10^{-6}$ ($V = 249 \pm 14$ neV to $V = 271 \pm 13$ neV) range \cite{atchison2006diamondlike}. Thus, reliable and reproducible DLC coatings are of great interest to experiments transporting and storing UCNs. Specifically, the EDM cell electrodes which apply large electric fields across a neutron storage volume must have excellent coatings, as neutrons will interact with them tens of thousands of times during measurement cycles. Using PENTrack \cite{schreyer2017pentrack} Monte Carlo simulations of UCN transport for the nEDM experiment at TRIUMF, it has been found that coating the EDM cell electrodes with DLC is crucial, leading to a 40\% to 100\% increase in the UCN counts detected \cite{SidhuPhD}, depending on the assumptions used in determining $W$ for NiP and DLC. A comparison of NiP with potential DLC properties is shown in table \ref{tab:materials}.  \\

\begin{table}[h!]
\centering
\begin{tabular}{|l|l|l|} \hline
Material & V (neV)     & $\eta$ = W/V (10$^{-4}$) \\ \hline 
DLC      & 220 - 271   & 0.01 - 0.5               \\
NiP      & 214 $\pm$ 5 & 3.5 $\pm$ 0.5   \\\hline        
\end{tabular}
\caption{Possible DLC coating properties compared to the baseline properties of electro-less NiP that would be used in the experiments at TRIUMF \cite{AKATSUKA2023168106}.}
\label{tab:materials}
\end{table}
Interest in using DLC films for UCN began in the late 1990's because of its high density (and therefore high real neutron-optical potential), low neutron absorption cross section, and expected low depolarization per bounce - making it equally useful for experiments with polarized and non-polarized UCN. It was hoped that DLC could replace the beryllium coatings used in UCN experiments \cite{atchison2006diamondlike}. To be used for UCN storage, DLC films need to have a high density, high purity, and a thickness greater than 100 nm \cite{mammei2010}. The thickness prevents UCN from tunneling through to the substrate material, which typically has much higher UCN losses \cite{Gol91}. Films must have very low levels of impurities that absorb or up-scatter UCN such as $^1$H or $^{10}$B \cite{PhysRevLett.39.1509}. We aim to develop a DLC coating with a real neutron-optical potential much higher than that of NiP, with a small imaginary component. A pure diamond would have a neutron-optical potential of 304 neV, setting the upper limit. \\ 

Along with its attractive properties for UCN storage and transport, DLC coatings also find many other practical applications due to its hardness, low friction coefficient and bio-compatibility \cite{bewilogua2014DLChistory}. They work well for components which require low wear in air, in vacuum, or in cryogenic settings \cite{fan2015carbon, miyai2009mechanical, tyagi2019critical}. DLC is often used for automotive components, cutting and machining tools, medical devices and implants, and hard disk drives among other things \cite{Hauert2004DLC, Robertson2002DLC, https://doi.org/10.1002/ppap.200731001}.\\

\begin{figure}[H]
    \centering
    \includegraphics[width=.65\linewidth]{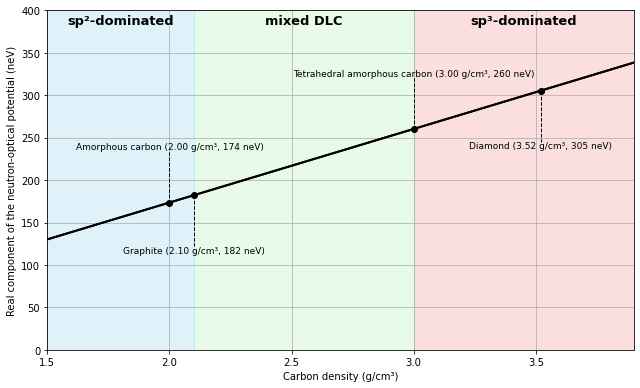}
    \caption{Fermi potentials in carbon films increase linearly with density. Blue region includes primarily sp$^2$ bonded films, green region includes films with near equal parts of sp$^2$ and sp$^3$ bonding, and the red region shows sp$^3$ dominated films. Higher Fermi potential results in fewer UCN losses due to material interactions.}
    \label{fig:density}
\end{figure}

DLC is an amorphous mixture of graphite and diamond forms of carbon, hybridized sp$^2$ and sp$^3$ bonds respectively. Its density, and thus neutron-optical potential are determined by the fraction of sp$^3$ (diamond) bonds. Diamond has a density of 3.52 g/cm$^3$ compared to  $\approx$ 2.1 g/cm$^3$ for graphite and thus a higher fraction of sp$^3$ bonds or higher density is desirable to make the best UCN guides. There are a variety of DLC types which cannot all be used for UCN applications, due to hydrogen contamination. These include amorphous carbon with mostly sp$^2$ graphite-like bonds (a-C), amorphous carbon with mostly sp$^2$ and hydrogen bonds (a-C:H), amorphous carbon with mostly sp$^3$ diamond-like bonds (ta-C), and tetrahedral amorphous carbon with mostly sp$^3$ and hydrogen bonds (ta-C:H). ta-C is preferred for UCN applications, due to its high density and minimal hydrogen contamination. A plot of neutron-optical potential versus carbon density is shown in figure \ref{fig:density}, with indications of different DLC bonding structures and their corresponding optical potential\\

While chemical vapour deposition can be used to create DLC films, in this work the films are produced via a plume of energetic carbon atoms resulting from the irradiation of a target with a pulsed laser. There are several models that describe DLC film formation. The subplantation model is used to describe the deposition of ta-C. In this model, sp$^3$ bonds are formed when C$^+$ ions or energetic neutral atoms originating from the carbon plasma plume or from surface atoms energized by plume collisions penetrate the growing film as discussed in section \ref{sec:DLC}. This mechanism, often referred to as the "knock-on" process, is characteristic of ion-assisted deposition \cite{Robertson2002DLC}. Initially, an amorphous sp$^2$ carbon film forms on the substrate. As the deposition continues, incoming carbon atoms with sufficient energy can penetrate the surface and become embedded within the bulk of the film. Those lacking the energy are deposited on the surface and contribute to increasing the film's thickness. During penetration, about 30\% of the atom’s energy is expended displacing other atoms, while the remaining energy is dissipated as heat, creating a localized thermal spike. This thermal spike leads to local densification of the film and promotes the formation of sp³ bonds. However, the same localized heat can also enable thermally activated diffusion, allowing some sp$^3$ bonded atoms to relax and migrate back toward the surface. As a result, the film grows through a balance of atoms adhering to the surface and those subplanting into the structure, leading to both densification and outward atomic movement. Experimentally, films with the highest sp$^3$ content are observed when the energy of the C$^+$ ions in the plume are around 100 eV \cite{Robertson2002DLC}. Below 100 eV, atoms do not have enough energy for the subplantation process to occur, and above 100 eV the thermal relaxation process dominates as the film is heated further resulting in sp$^3$ fraction decrease. This model predicts a maximum sp$^3$ fraction of 0.8 \cite{Robertson2002DLC}. It has been shown that deep cryogenic treatment of DLC coatings can further increase the sp$^3$ fraction \cite{PENG2022109189}. \\

At The University of Winnipeg (U.Winnipeg), a new thin films coating facility has been built over 2023-2025: the UCN guide coating facility (GCF). The GCF employs pulsed laser deposition (PLD) to apply UCN-reflective coatings to the inside of cylindrical UCN guides, and other experimental components. The major components of the facility were purchased from Virginia Tech which coated guides for the UCN source and UCNA experiment at Los Alamos National Laboratory and the UCN source at the PULSTAR reactor \cite{lanlucnsource,ucnapub2013, KOROBKINA2014169, UCNA}. We note that while at Virginia Tech, DLC films were made with a neutron optical potential of $>$260 neV\cite{MakelaPhD} on quartz tubes with hydrogen content of less than 1\% \cite{mammei2010}. We continue this development and in this paper we describe the facility at U.Winnipeg and first coating samples.

\section{The Facility}
\begin{figure}[H]
    \centering
    \includegraphics[width=1.0\linewidth]{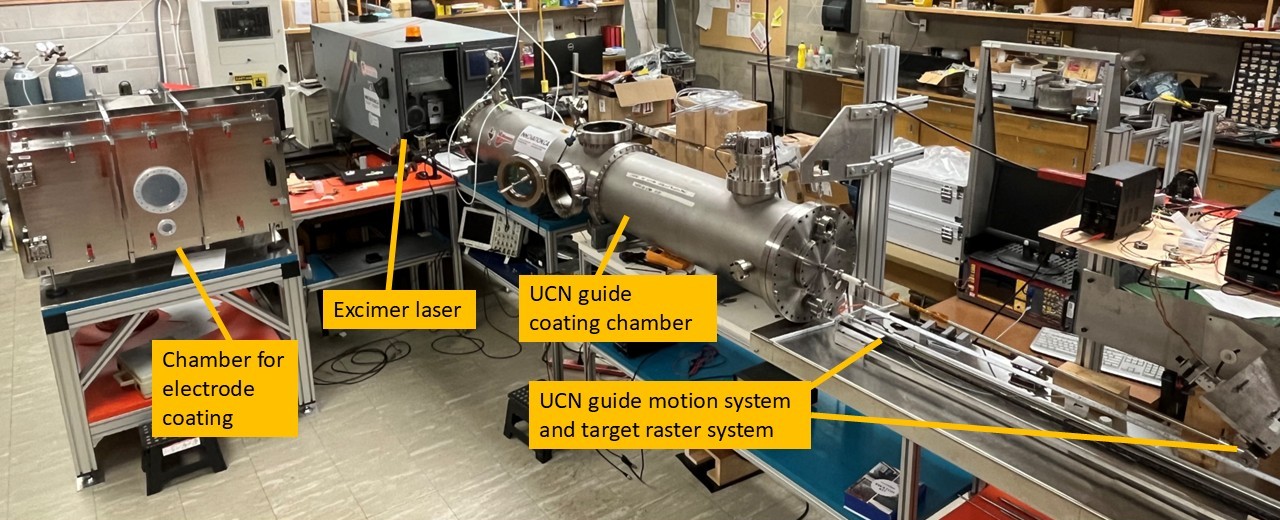}
    \caption{Annotated photograph of the GCF with major components labeled.}
    \label{fig:schematic}
\end{figure}

The GCF uses thin film deposition techniques to coat UCN transport and storage components for the TUCAN experiment. An annotated picture of the main infrastructure is shown in figure \ref{fig:schematic}. The facility uses an excimer laser (248 nm, 850 mJ/pulse) to make coatings on components enclosed inside a 3 m long, 14" diameter cylindrical vacuum chamber. This chamber can provide a coating onto the inside of UCN guides up to 1 m long with a 200 mm outer diameter. Table \ref{tab:chamber} lists the sizes of various geometry of parts that have been coated in the cylindrical chamber which is discussed in \ref{sec:coatingchamber}.  An additional vacuum chamber currently being commissioned will allow for larger flat geometries to be coated as well, namely the circular TUCAN EDM cell electrodes. The facility is currently focused on PLD of DLC, but can also be used to coat substances like isotopically enriched $^{58}$Ni using e-beam evaporation \cite{mammei2010}. After coating, thicknesses are measured using a Dektak$^3$ profilometer before samples are sent to other institutions for further analysis of their diamond structure and optical properties using a variety of surface science techniques.

\begin{table}[]
    \centering
    \caption{Dimensional parameters of the items that can be coated in the cylindrical deposition chamber.} 
  \label{tab:chamber}
    \begin{tabular}{|c|c|c|} \hline
        \textbf{Geometries} & \textbf{Diameter (mm)} & \textbf{Length (mm)}  \\\hline
         Inside of Tubes & 25 - 200 & < 1000 \\ \hline
         Outside of Rods & 5 - 50 &  < 1000 \\ \hline
         Face of Discs & 150 & <50 \\ \hline
         Plates & width < 50  & <1000 \\ \hline
    \end{tabular}
    
\end{table}

\subsection{Pulsed Laser Deposition}\label{sec:DLC}

The process of laser ablating a target occurs in three main steps: laser induced plume production, laser-plume interaction, and the plume evolution. When the laser strikes the target, local extreme heating vaporizes the material into a plasma, which is then ejected in a plume. The energy of the plasma plume can be increased by the absorption of laser energy after being created, or by increasing the laser energy which causes target ions to have higher kinetic energy when being emitted from the target. After the laser pulse, the plume expands outwards in the vacuum and many processes such as momentum transfer, ionization, recombination, and radiative de-excitation occur within the plume. These all affect the plume energy and species which would eventually deposit on the substrate. \\

The laser energy must be above the energy threshold for vaporization of the material for a plasma to be created, 33 eV for graphite \cite{MONTET196719}. For pulse lengths longer than a few picoseconds, heat diffusion and optical absorption affect how energy is distributed inside the target material. This can cause large atomic clumps of target material to be ejected. The laser wavelength, energy, and pulse length must be carefully tuned to create the desired plasma plume. By using 25 ns-long pulses, the plasma plume forms while the laser is still striking the target. As the plume grows it becomes further ionized by the laser. Using this technique plasma plumes with energies of the order of 100s of eV can be produced \cite{RevModPhys.72.315}. \\

\subsection{Laser}
A Lambda Physik LPX305i excimer laser (1997 manufacturing date), with a 50 L Nova lasing tube provides 248 nm light with a 25 ns pulse width for the DLC coatings at the GCF. The laser uses a mixture of ultra high purity grade Kr, F (5\%)/He (balance), and Ne (buffer) gases. The maximum laser pulse energy at 248 nm is 850 mJ/pulse at 15 Hz and has a rectangular beam shape of $\sim$10 mm by 30 mm before it is collimated. Special laser windows have been installed where the back reflecting laser window is a planar mirror while the front laser window is slightly curved, providing some reduction in laser beam divergence compared to a planar mirror. This was experimentally found to provide the smallest focused beam spots on the ablation target.


\subsubsection{Optics}
\begin{figure}[H]
    \centering
    \includegraphics[width=.65\linewidth]{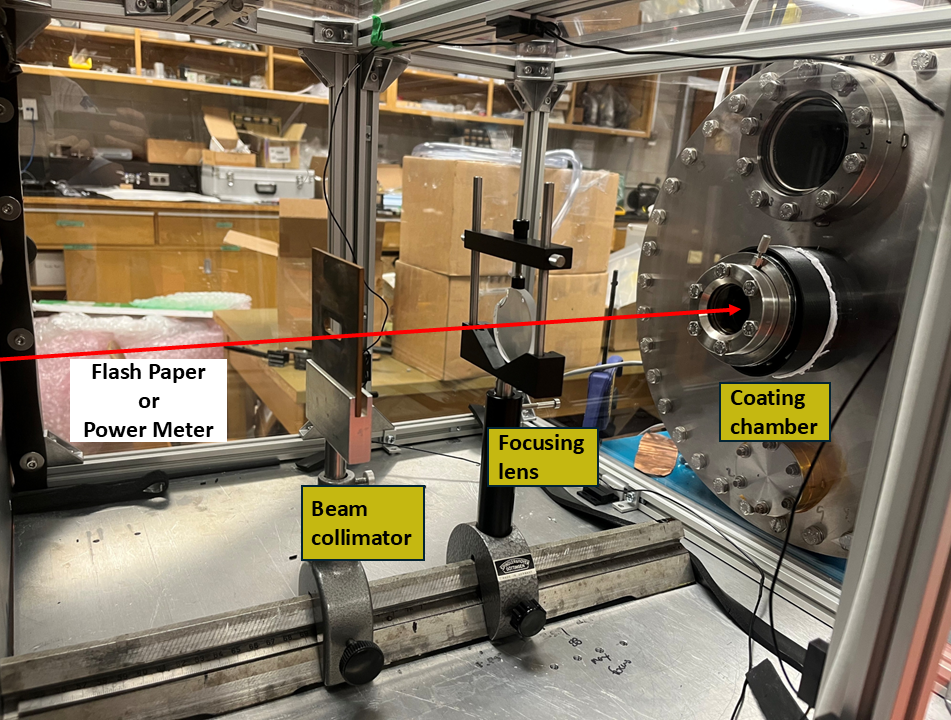}
    \caption{Left to Right: Laser (out of frame), collimator, focusing lens, anti-reflective window, coating chamber. Lens is 2" diameter for scale. Laser path shown by red arrow.}
    \label{fig:optics}
\end{figure}

After exiting the laser tube a copper collimator with a rectangular cross section of 9 mm by 30 mm is placed in the laser path 10 cm away from the laser exit. This blocks parts of the halo around the central 248 nm beam from reaching the target and reduces back-reflections to the laser. To ensure the internal optics of the laser are aligned, the beam shape is monitored by burning images of single laser pulses onto thermal receipt paper. The position of the paper is shown in figure \ref{fig:optics}. The energy after the collimator is monitored using a handheld Gentec EO laser power meter periodically inserted into the laser beam. After the collimator, the beam energy is reduced to several hundred mJ/pulse (depending on laser high voltage settings). After collimating, a 1.0 m focal length 248 nm anti-reflection coated fused silica plano-convex lens from CVI Optics (PLCX-50.8-749.5-UV-248) focuses the laser beam. The beam then travels through an anti-reflection coated vacuum window (also from CVI optics, Part no. W1-PW1-2025-UV-248). Typically, only the air side of the window has this coating, as the vacuum side gets coated with the PLD material, meaning an anti-reflective coating on the inside of the chamber would quickly become occluded. The optics between the laser and vacuum chamber can be seen in figure \ref{fig:optics}. \\

\subsubsection{Laser Safety Enclosure}
A polycarbonate/aluminum extrusion safety enclosure was built to fit around the optical area of the facility. The enclosure and several coating chamber viewing ports are interlocked to prevent the laser from firing if not properly positioned. This allows the class IV laser system to be reclassified as a class I laser system reducing the safety requirements. \\
\subsection{Coating Chamber}
\label{sec:coatingchamber}
\begin{figure}[H]
    \centering
    \includegraphics[width=1.0\linewidth]{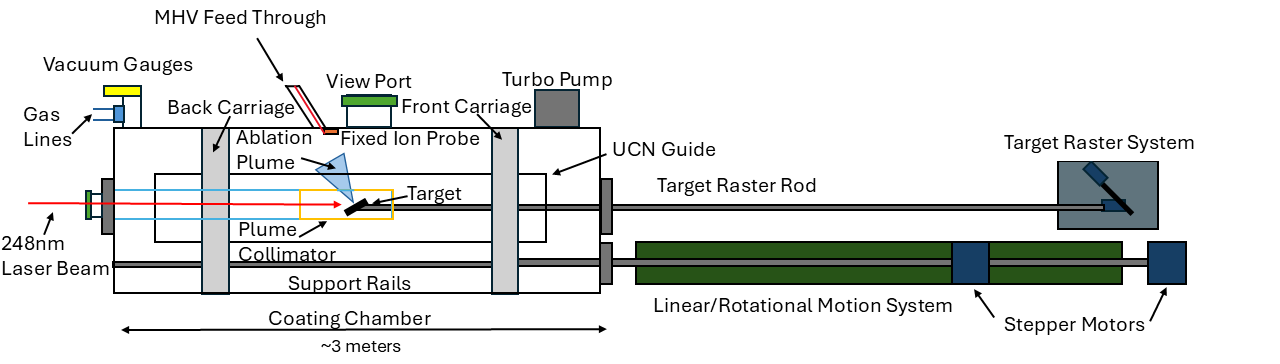}
    \caption{Schematic diagram of the coating chamber used to apply thin films coatings onto the inside of UCN guides. The UCN guide tube is loaded into two carriages that rotate and translate the tube over the plasma plume.  The rotation and translation motion is controlled outside the vacuum chamber.  The PLD target rastering is also controlled outside the vacuum chamber.  }
    \label{fig:chamberschematic}
\end{figure}

\begin{figure}[H]
    \centering
    \includegraphics[width=0.34\linewidth] {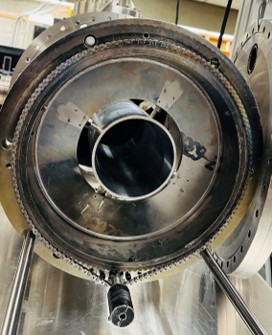} \hspace{2mm}
    \includegraphics[width=0.4\linewidth]{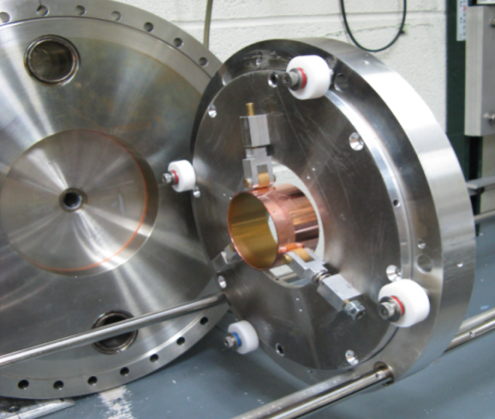}
    \caption{Left: front carriage showing black helical rod coupling and chain/sprocket system. Right: back carriage showing UCN guide tube mounting plates holding a copper UCN guide sample tube. }
    \label{fig:carriages}
\end{figure}

\begin{figure}
    \centering
    \includegraphics[width=0.45\linewidth]{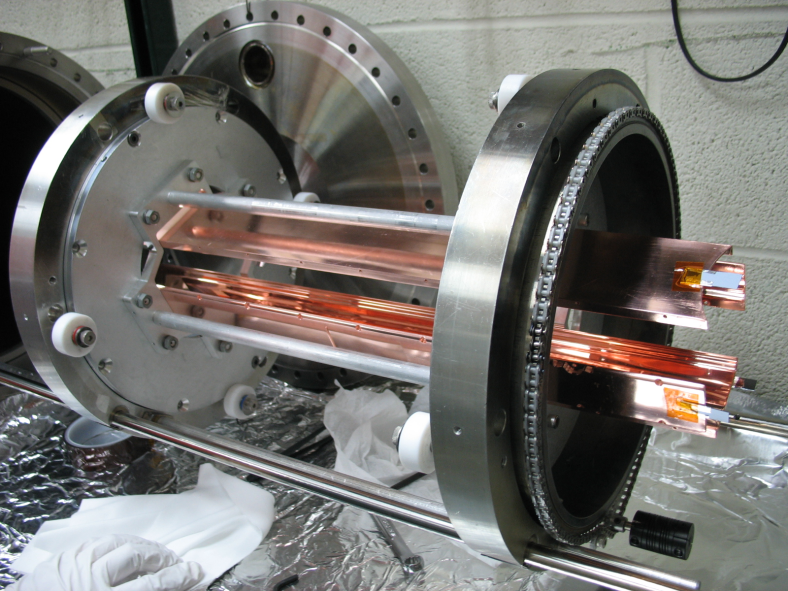}
    \hspace{2mm}
    \includegraphics[width=0.47\linewidth]{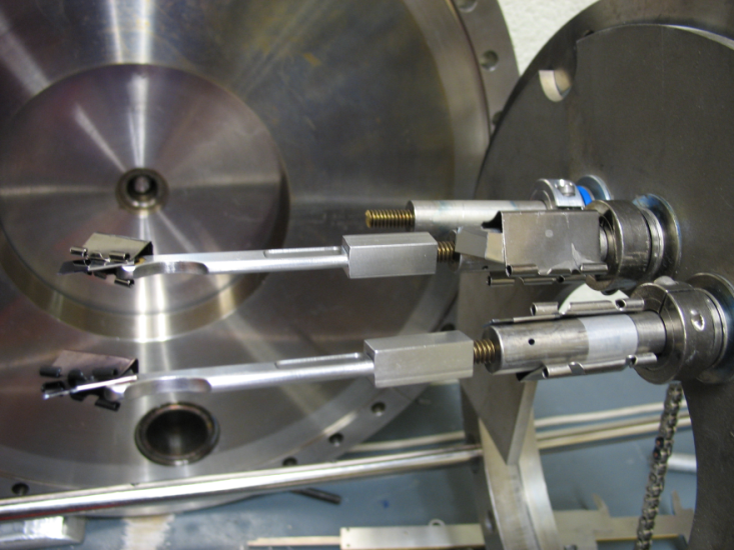}
    \caption{Example of various carriage mounts that allow the coating of long plates (left) and rods (right). On the left photograph, one can see small strips of silicon wafer, our ``witness strips'', held on to the copper plates via orange colored Kapton tape. }
    \label{fig:parts]}
\end{figure}

The facility uses a 3 m long, 14" diameter 304L stainless steel vacuum chamber, pumped out by a 300 L/s (N$_2$) Varian Navigator 301 turbo pump. Coating depositions are typically made under holding vacuum in the region of 10$^{-6}$ mbar. As shown in figure \ref{fig:chamberschematic}, the ablation target, described below, is held in the center of the chamber by a rod on one end of the chamber, while the laser beam enters from the other end. The UCN guide tube to be coated is mounted in a dual carriage system, shown in figures \ref{fig:carriages} and \ref{fig:parts]} that allows the tubes to be rotated and linearly translated over the target, target rod, and laser beam path. Each carriage is supported on a dual linear rail system with linear bearings and are connected to each other with aluminum support rods. A 10" diameter thin section bearing is fixed to each carriage and allows the rotation action. Mounting plates are fixed to this bearing which have screws that center and secure the UCN guide tube in the center of the chamber. Torque from the 1/4" diameter drive rod is coupled through a helical rod coupling to rotate a sprocket chain (ANSI 25) driven rotation system mounted to the front carriage's 10" diameter bearing. This allows torque to be transferred from outside the chamber through the 1/4" diameter drive rod, to the chain mechanism, and finally to the UCN guide tube. The torque is further transferred to the back carriage, shown in figure \ref{fig:carriages}, through the guide tube which has the same 10" diameter bearing and ucn guide tube mounting plates. The 1/4" rod penetrates the vacuum via an oring sealed quick-connect fitting and is coupled to a custom dual stepper motor rotation/linear motion system, shown in figure \ref{fig:schematic}. The end of the drive rod is connected to the "rotation" stepper motor which itself is fixed to a linear drive system, controlled by another stepper motor. Linear drive speeds are typically 2.6 cm/min and UCN guide tube rotation speeds are 12 rpm. Figure \ref{fig:parts]} shows some photographs of how other geometries beside tubes are spun over the plasma plume. Detailed information concerning the carriage system and the various geometries that have been coated is found in \cite{mammei2010}.

\subsubsection{Target}
\begin{figure}[H]
    \centering
    \includegraphics[width=.8\linewidth]{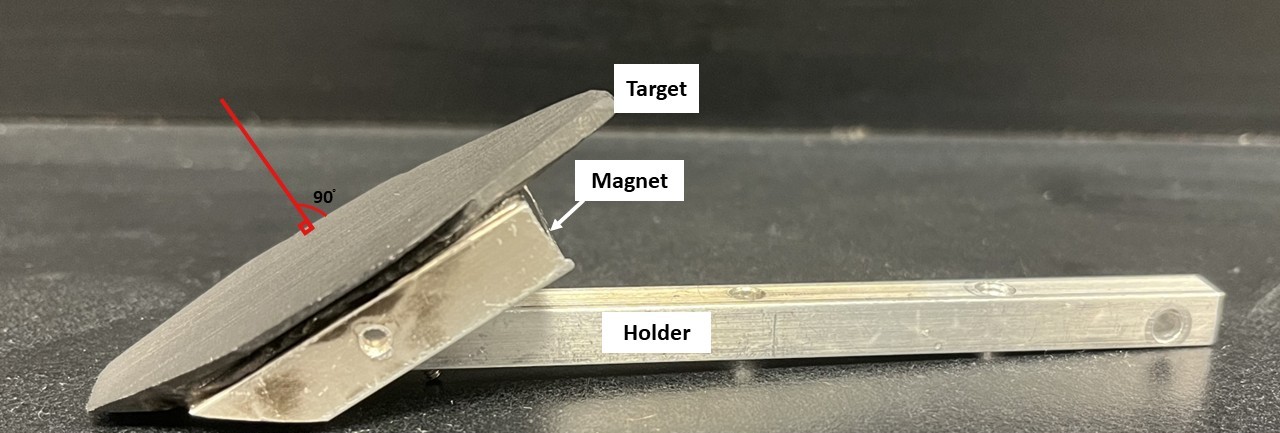}
    \caption{Side-view of target holder with a target mounted. The target is a 29 x 58 x 6 mm octagonal piece of pyrolitic graphite, with a 37.5 x 19 x6 mm N52 neodymium magnet glued to the back and held by a custom machined aluminum holder. Angle shown to demonstrate how angles are measured in section \ref{sec:TOF}.}
    \label{fig:target}
\end{figure}

The facility uses an ablation target which fits inside of the coating chamber and is placed at the center. The target holder (Figure \ref{fig:target}) aligns the target at an angle of 30$^{\circ}$ relative to the laser beam, as it was found to produce the best coatings \cite{mammei2010}. The target is a 29 x 58 x 6 mm octagonal piece of pyrolytic graphite (either substrate nucleated or continuously nucleated) purchased from \cite{MineralsTech_PyrolyticGraphite_Properties}. Highly oriented pyrolytic graphite is used due to its high purity (99.995\%), thermal conductivity, and surface uniformity \cite{WasyZia2023}. Fixed to the back of the graphite target is a 37.5 x 6 x 19 mm N52 Neodymium magnet with a field in the direction of its thickness. The magnet is glued to the target using Hysol 1C high vacuum rated epoxy. The magnetic field, measured to be 200 mT at the surface of the graphite, increases the number of collisions the plasma plume species have with each other. This breaks up larger clusters which could help to make a smoother and more diamond-like film \cite{Hou1998InfluenceOA}.


\begin{figure}[H]
    \centering
    \includegraphics[width=1.0\linewidth]{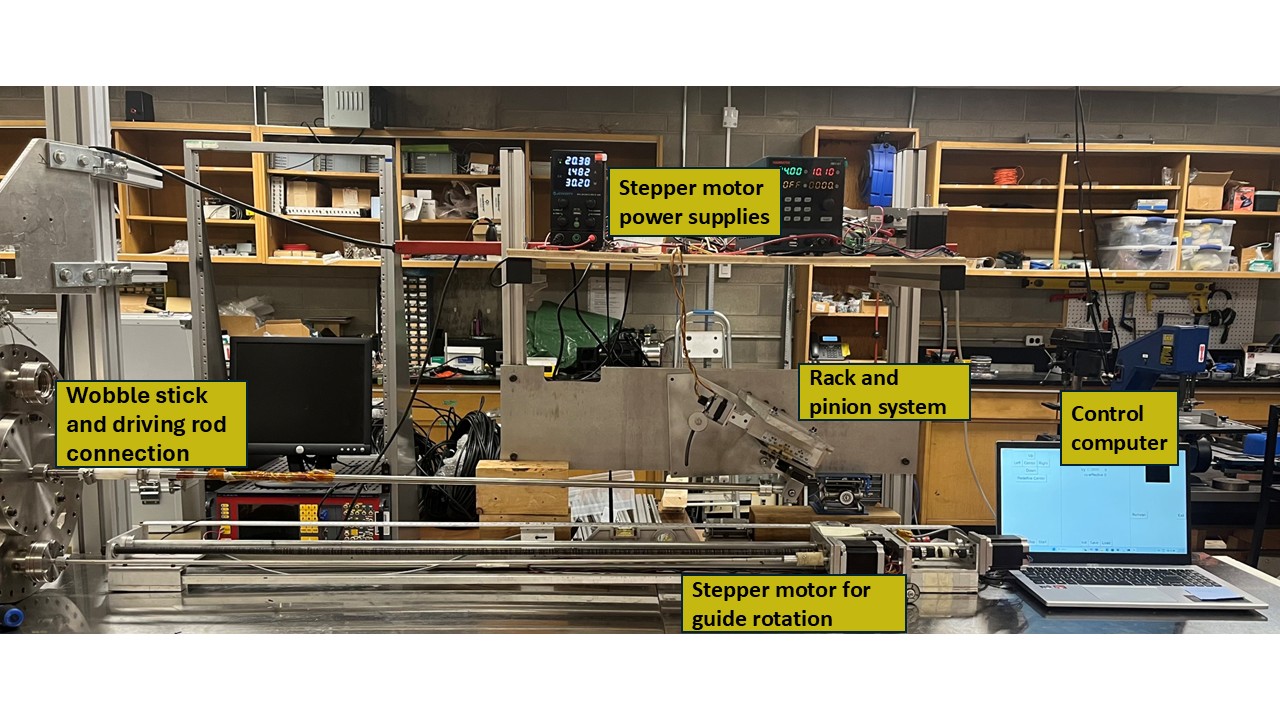}
    \caption{Guide rotation and target motion system. The guides are rotated during film deposition by a stepper motor located (bottom right) to ensure coating uniformity. The graphite target may be moved both horizontally on-axis with the chamber and vertically via the labeled rack and pinon system and a linear actuator.}
    \label{fig:motion}
\end{figure}

The target and magnet are positioned in the chamber at the focal point of the laser system to maximize the on target laser energy density. The target is isolated from ground so that an additional bias voltage of hundreds of volts can be applied to the target. The target is negatively biased rather than the substrate, as applying a voltage to the rotating and translating substrate is challenging. It was found experimentally, that a negative 100 to 200 V target bias improved adhesion of the DLC on the copper UCN guide substrates used in the UCNA experiment\cite{mammei2010}. The bias on the target has an opposite polarity to what usually  applies to the substrate in a typical PLD setup. The custom machined target holder is depicted in Figure \ref{fig:target}. This holder attaches to the target magnet and mounts into an aluminum wobble stick which extends to the far end of the vacuum chamber. The rod then passes a Quick Connect Viton O-ring seal, which acts as an electrical insulator. On the air-side of the rod, another aluminum rod is clamped on using a nylon clamp, ensuring the rod inside the chamber is electrically floating. This rod is connected to a two-axis motion system. Using the motion system shown in Figure \ref{fig:motion}, the target is moved vertically and horizontally inside the chamber. Horizontal motion is provided by an on axis, rack and pinion system. Vertical motion is provided by an actuated linear slide bearing tilted at an angle to match the target angle of 30$^\circ$. The raster pattern is an ellipse with a 35 mm major axis (along the length of the target) and 20 mm minor axis (along the width of the target). The minor axis motion speed is ten times that of the major axis speed and it takes about 3 minutes for the laser to draw a full ellipse on the target. This motion prevents the target from having holes drilled through it and allows for better control of what sections of the ablation plume hit the substrate \cite{mammei2010}. This is also managed using another, different collimator, which is further discussed in section \ref{sec:collimator}. By rastering the target, the laser spot and collimator are able to stay fixed during coatings. \\

Coating substrates are loaded first in the carriage system and pushed all the way to left of the coating chamber as shown in Figure \ref{fig:chamberschematic}. Then the target is attached to one end of the  wobble stick and loaded into the chamber and connected to the raster motion system described above. After targets have been machined, they are wiped clean using ethanol, and ultrasonic cleaned in de-ionized water for 15 minutes, rinsed in DI water, and left to air dry. The target is always ``laser cleaned'', having the laser ablate material from the full target raster ellipse, before allowing the substrate to pass over the laser ablation plume. This prevents target surface contaminates from being part of the coating. After ablating a target, it must be cleaned again to have a flat surface for ablation. The target is sanded flat using 800 grit SiC sandpaper, and rinsed and ultrasonic cleaned again. A target can be re-used and cleaned until it is too thin to provide enough carbon for a coating and it must be remade.





\subsection{Plume Collimator \label{sec:collimator}}
\begin{figure}[h!]
    \centering
    \includegraphics[height=8cm]{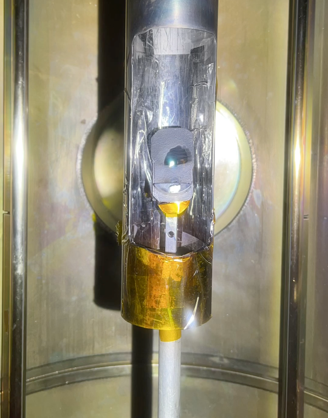} \hspace{2mm}
    \includegraphics[height=8cm]
    {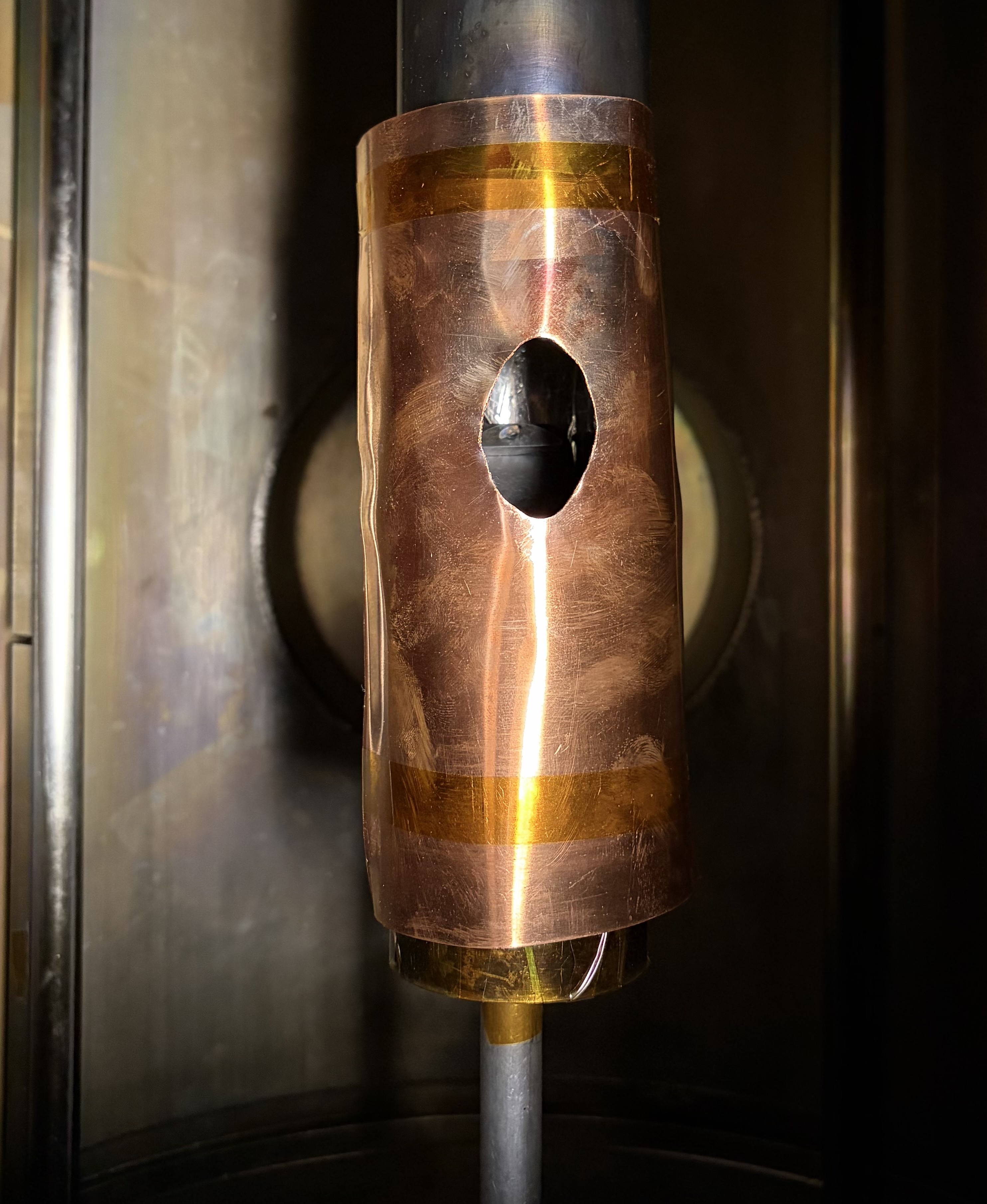}
    
    \caption{Top view of the target, without (left) and with (right) the Cu collimator. A blackened elliptical area is beginning to form on the graphite target on the top figure as a result of the target raster system. Also the target mounting rod is seen at the bottom of both photographs.}
    \label{fig:collimator}
\end{figure}

As discussed in section \ref{sec:intro}, it has been observed that a C$^{+}$ ion energy of 100~eV depositing on the substrate leads to the densest DLC films \cite{Robertson2002DLC}. Therefore, it is of interest to prevent carbon ions with energies that greatly differ from 100 eV from depositing on the substrate. A collimator is placed around the ablation target to control the energy of the plume that hits the substrate, since the center of the plume contains the highest energy and smallest cluster sizes \cite{RevModPhys.72.315}. However, a narrow collimator means deposition rates are low and coatings take longer, possibly needing to be monitored overnight to maintain laser energy. To determine which parts of the plume must be collimated, an in-situ diagnostic tool, an ion probe, was swept across the plasma plume during ablation to characterize its energy. The collimator shown from above the target in figure \ref{fig:collimator} is designed using measurements with the ion probe discussed in the following section.   \\

\begin{figure}[h!]
    \centering
\includegraphics[width=.8\linewidth]{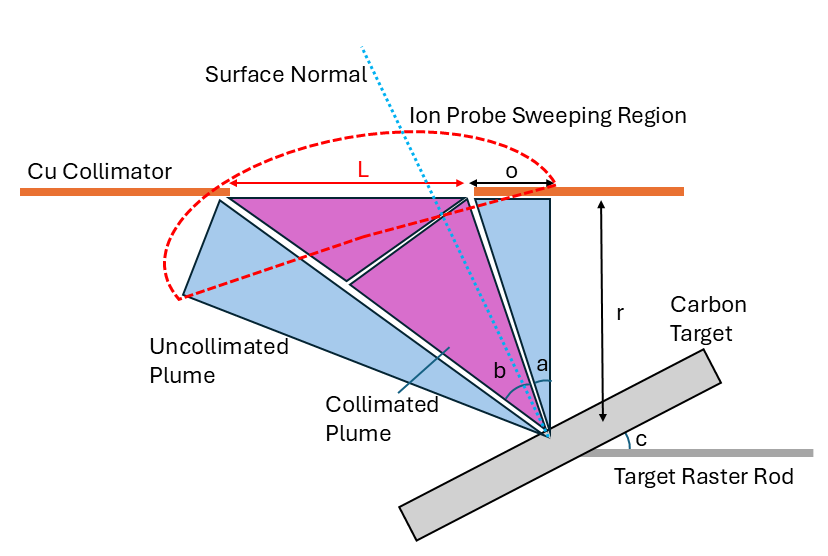}
    \caption{\textit{Side view.} Design of plasma plume collimator, used to exclude non-100 eV particles from hitting the coating substrate. The carbon target is placed at a tilt angle $c$ =  30$^{\circ}$. The desired collimation angle $b$ = 40$^{\circ}$ creates an angle $a$ = 20$^{\circ}$ from vertical, which using the measured vertical offset of the collimator $r$ = 25 mm gives a distance $o$ = 9 mm to place the collimator from the target center. The length $L$ needed to collimator the plasma plume to 100 eV is then calculated to be 32 mm.}
    \label{fig:collimator_design}
\end{figure}

The aperture on the plume collimator was designed using measurements of the ion probe shown in figure \ref{fig:probe_data}. Data taken without the collimator informed the design shown in figure \ref{fig:collimator_design}. After sweeping the ion probe through the region shown inside the dashed red line, the angle, $b$, was determined to be 40$^{\circ}$ (containing angles 60-100$^{\circ}$, taking 90$^{\circ}$ as surface normal). The angle, $a$, from vertical, determined the offset distance $o$, that the collimator should be placed relative to the target center (along its length). Measuring the distance from target center to collimator, $r$, determines the length $L$ needed for the collimator opening to let the 100 eV (purple) part of the plume through to the substrate, blocking the plume with energies differing from 100 eV (blue). Specific values for the measurements are noted in the caption of figure \ref{fig:collimator_design}. The final result is the collimator shown in figure \ref{fig:collimator}, which is a oval of major axis length 32 mm and minor axis length 22 mm.

\subsection{TOF Ion Probe}\label{sec:TOF}
\begin{figure}[H]
    \centering
    \includegraphics[width=0.4\linewidth]{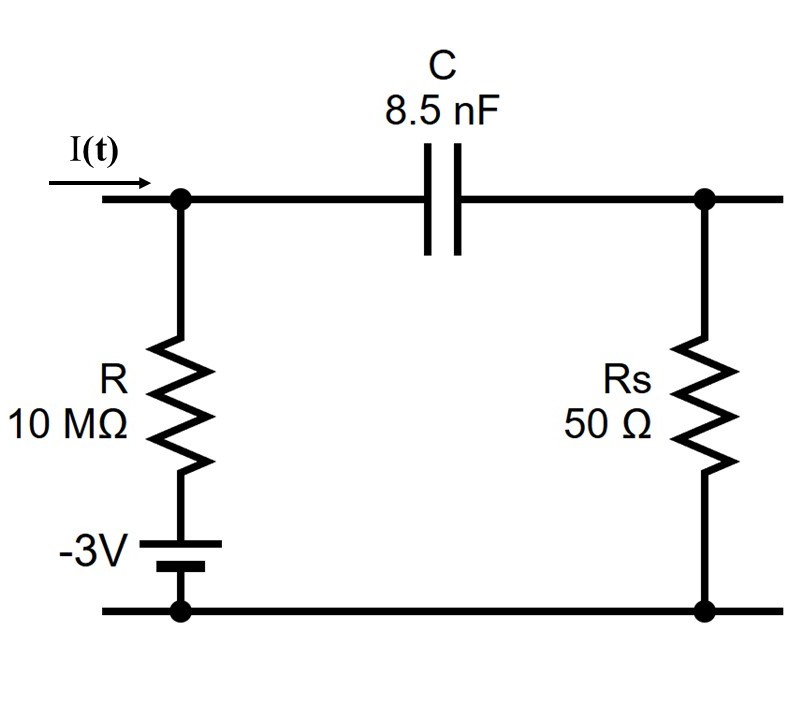}
    \caption{TOF ion probe circuit, where R = 100 $\Omega$ is the resistor, C = 8.5 nF is the capacitor and R$_s$ = 50 $\Omega$ is the impedance of the oscilloscope. I(t) represents the ion current created by ions striking the copper tip of the ion probe.}
    \label{fig:circuit}
\end{figure}

The time-of-flight (TOF) ion probe, or Langmuir probe, is an essential tool for monitoring the plasma plume energy during the PLD process \cite{2003JAP....93.3627H, Langmuir_probe, Langmuir_Probe_Limitations}. In our case, the probe is used to measure the TOF of the carbon ablation plume, which we assume to be primarily composed of C$^+$. The facility uses two different kinds of ion probe in its operation: one fixed, and one that can sweep over the plasma plume. The fixed probe uses a small copper disc located at 90$^{\circ}$ relative to the target plane. This is fixed with kapton tape to the top of the chamber and connected to an insulated vacuum wire that exits the chamber through an MHV electrical feed-through. The circuit (figure \ref{fig:circuit}) connected to the ion probe is a biased RC circuit with R = 10 M$\Omega$ and C = 8.5 nF. As ions bombard the copper tip, a current I(t) is generated and the corresponding voltage is read across an oscilloscope with 50 $\Omega$ impedance.

\begin{figure}[H]
    \centering
    \includegraphics[width=0.6\linewidth]{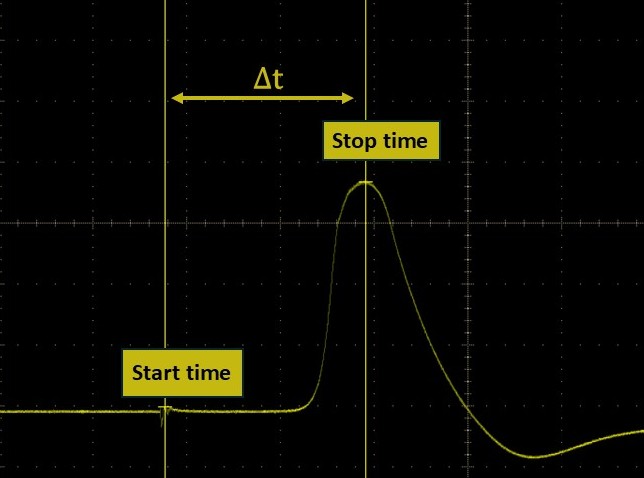}
    \caption{Oscilloscope trace of an example ion probe signal. Start and stop time of TOF labeled. TOF $\Delta$t = 1.7 $\mu$s. Vertical scale of 200 mV/division, horizontal scale of 800 ns/division. Kinetic energy of (53 $\pm$ 10) eV is calculated for this pulse.}
    \label{fig:scope_trace}
\end{figure}

The kinetic energy is derived from the measured speed and the mass of the assumed ion species. The speed is calculated via measuring the distance between target and probe and measuring the time of flight of the ions. To measure a time-of-flight, a UV photodiode is used for the start time. The peak of the positive ion voltage signal as seen on the oscilloscope in figure \ref{fig:scope_trace} is used as the stop time. By measuring the distance between target and probe tip, and time difference between the laser pulse and ion peak, a speed can be determined. Then assuming the dominant ion species is a single $^{12}$C$^+$ an average kinetic energy is calculated to be (53 $\pm$ 10) eV in figure \ref{fig:scope_trace}.  \\

\begin{figure}[H]
    \centering
\includegraphics[width=0.6\linewidth]{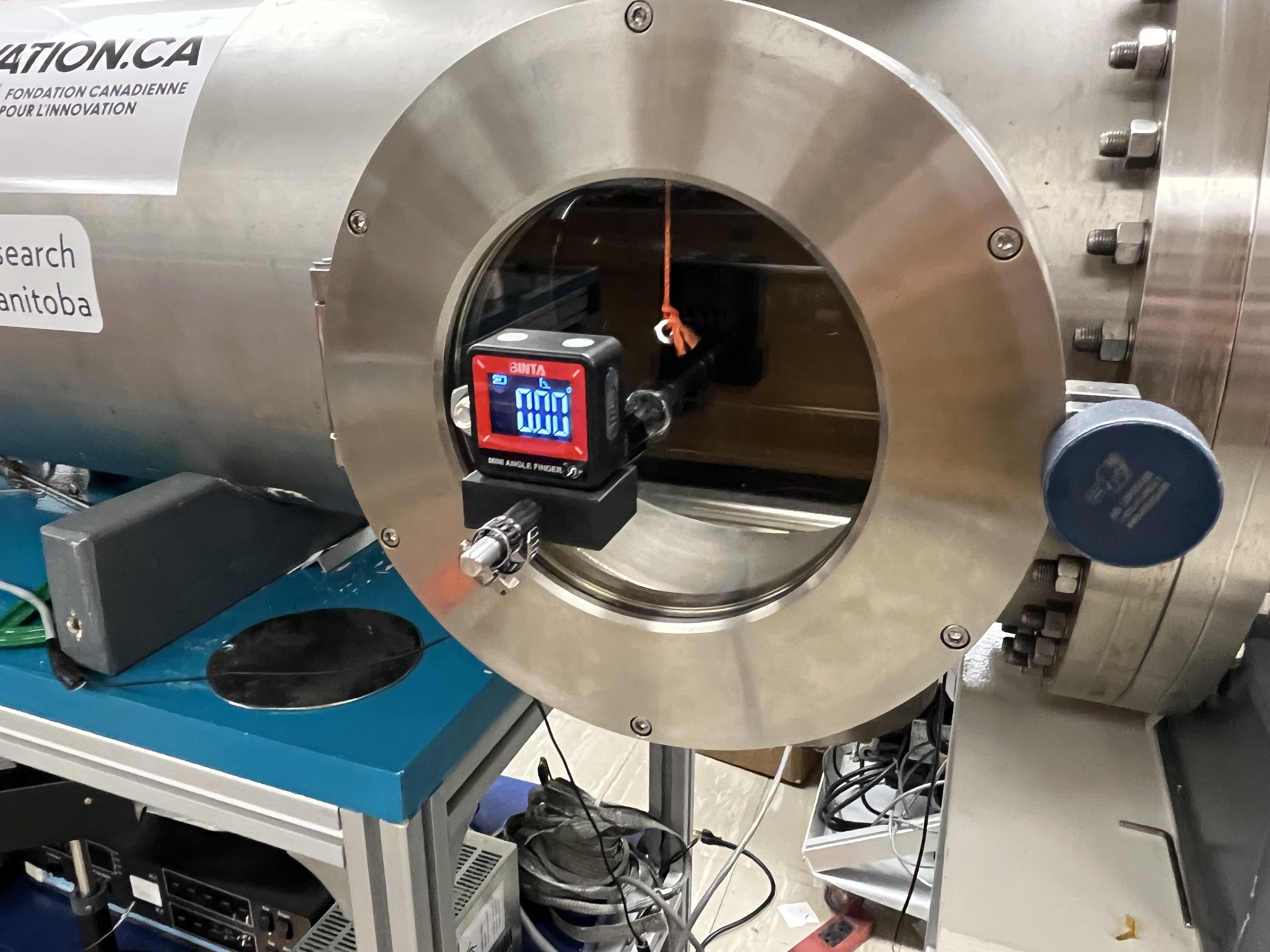}
\includegraphics[width = 0.6\linewidth]{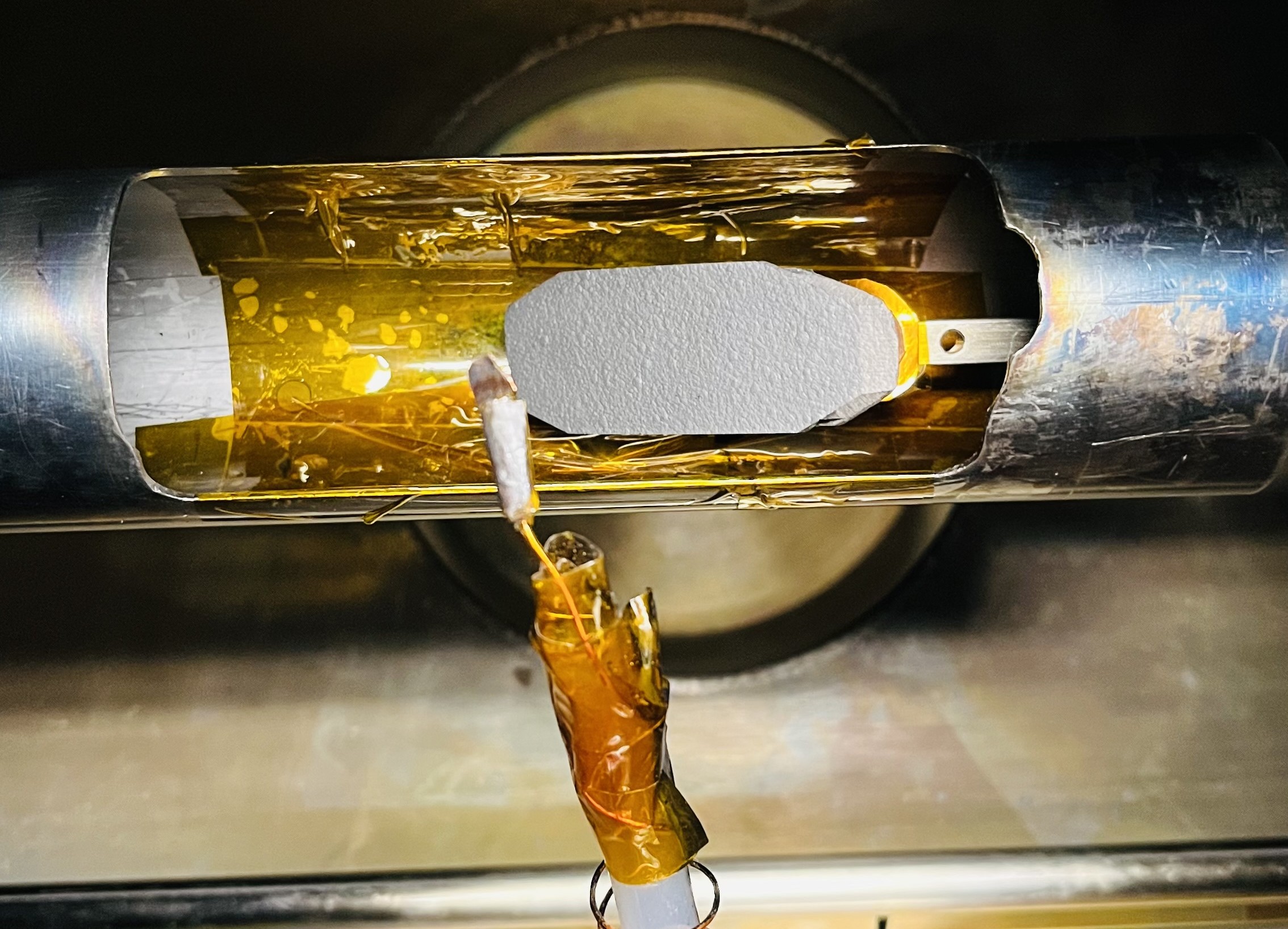}
    \caption{Top: Electronic inclinometer is placed on outer mount which rotates with the probe tip on the inside of the chamber. This allows for precise determination of the angle of the probe relative to the target normal. Bottom: aerial view of the movable ion probe tip above the 58 mm long octagonal graphite ablation target. Probe is swept left to right in this picture to sample the plasma plume.}
    \label{fig:ion_probe}
\end{figure}

\begin{figure}[H]
    \centering \includegraphics[width=.80\linewidth]{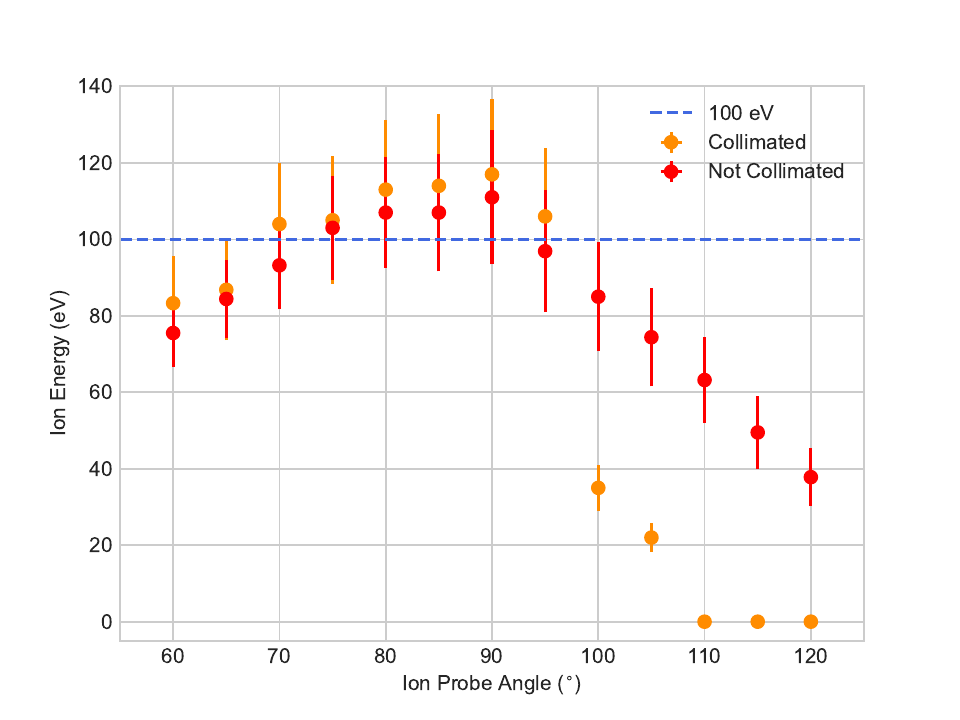}
    \caption{Plot of kinetic energy (eV) as measured by the ion probe versus angle relative to the plane of the ablation target. 90$^{\circ}$ is normal to the target plane (figure \ref{fig:target}). Data points from 70-95$^{\circ}$ agree with 100 eV goal, and other regions should be collimated further.}
    \label{fig:probe_data}
\end{figure}

The movable probe is used to inform the design of the plume collimator. This probe sticks through a vacuum window on the coating chamber, allowing it to be rotated above the center of the target, sweeping through a region shown in figure \ref{fig:collimator_design}. The angle relative to the target plane is measured by an inclinometer as seen in figure \ref{fig:ion_probe}. An example of a 90$^{\circ}$ angle is shown in figure \ref{fig:target}. After sweeping across the plasma plume and tuning the energy to be 100 eV near the target normal, an example plume collimator was made and placed over the ablation target. Figure \ref{fig:probe_data} shows the collected data. We see that we had some success collimating lower energy ions at angles greater than 90$^{\circ}$. We also see that data points at angles below 70$^{\circ}$ degrees do not agree with our goal of 100 eV, and should be collimated further in a next design.

We use the fixed ion probe to set the laser energy required to produce 100 eV C+ ions at the target's surface normal, and then monitor this energy after each full pass along the guide length (about 45 minutes for a 1 m long guide) throughout coatings. We have found that during long coatings, the laser energy must be increased as a build up of coating on the vacuum side of the anti-reflective window reduces the laser transmission through the window. Degradation of laser gasses also contributes to drops in laser power as well, as seen by laser energy measurements during coatings. Although it is extremely challenging to draw any plasma physics conclusions using the ion probe, it serves as an in-situ tool that provides basic information about the ablation plasma and helps make coatings at at a specific plume TOF consistently. 


\section{First DLC coatings}
\label{sec:firstcoatings}
\subsection{DLC coatings during Commissioning}
\begin{figure}[H]
    \centering
    \includegraphics[width=0.4\linewidth]{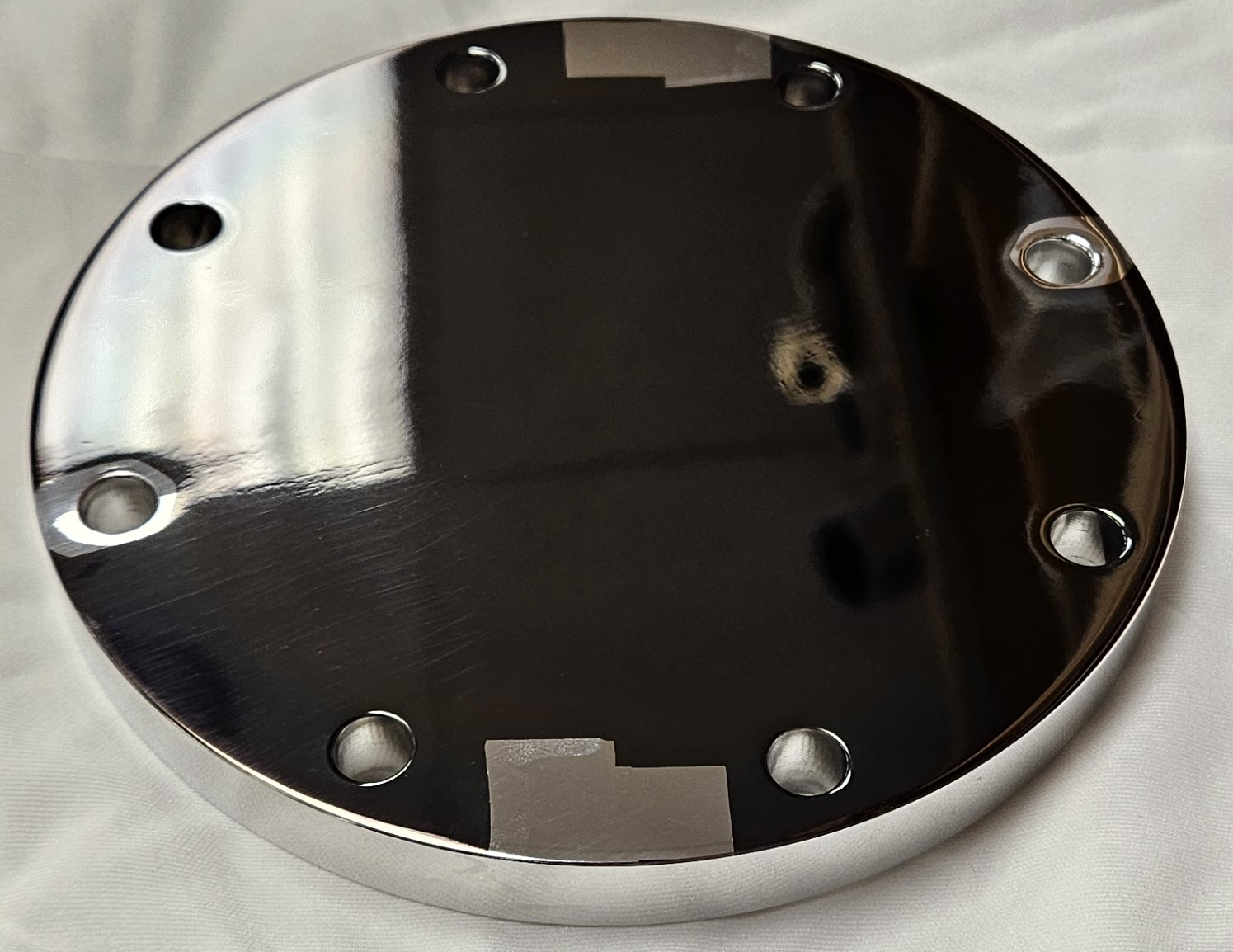}
    \hspace{2mm}
    \includegraphics[width = 0.33\linewidth]{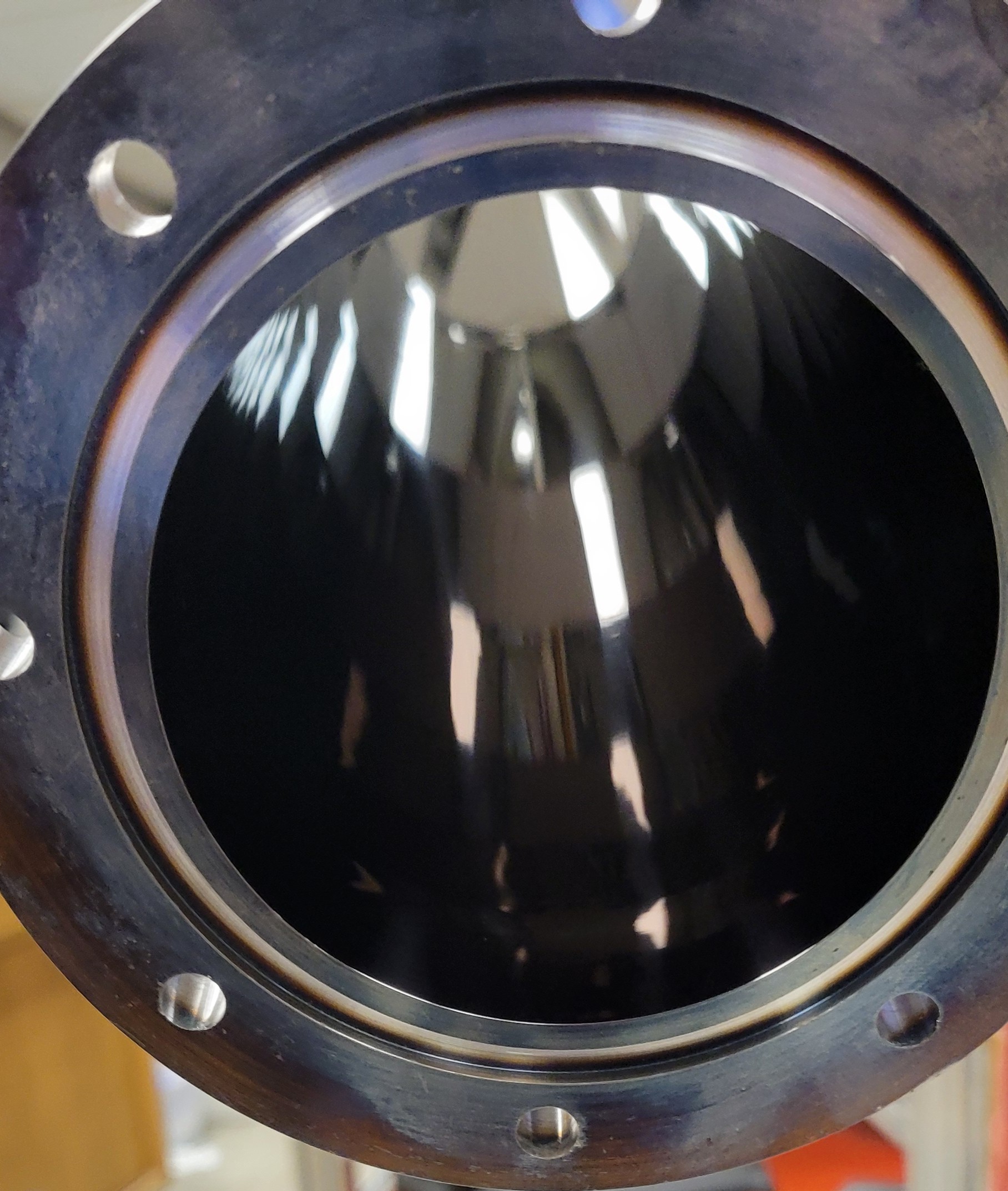}
    \caption{Top: picture of the DLC coated 150 mm diameter UCN-guide flange. Non-coated sections on top and bottom of flange were covered by Si witness strips used for profilometry and XRR measurements. Bottom: picture of the inner surface of the DLC coated 95 mm I.D. UCN-guide. Si witness strips are mounted to the outer-flange of the guide such that they do not cover the inner surface.}
    \label{fig:coating}
\end{figure}

During commissioning the facility coated several 100 (length) x 100 (outer diameter) x 2.5 mm wall thickness tube samples without a plume collimator or ion probe feedback. Several strips of cut silicon wafer, called witness strips, were fixed to the inside of the tubes as well as the outer diameter extending past the length of the tubes. In figure \ref{fig:parts]} (left) silicon witness strips can be seen protruding past the copper plates. It was found that if we employed a target bias then the coatings on the tube samples made this way would not delaminate when left sitting in air for 72 hours; tube samples that did not have a target bias had some delamination spots after sitting in air for 24 hours. After this, a first coating was completed on a 1 m long aluminum 100 mm OD x 2.5 mm wall TUCAN UCN guide along with a matching flange, both with a target bias of -200 V and no plume collimator. The primary motivation of these coatings were to test whether the facility met the mechanical requirements to make coatings with the appropriate thickness and adhesion properties for a full length guide and flange. The coated flange and guide are shown in figure \ref{fig:coating}. We note after a year, the coating has not delaminated from these parts. Collimating the plume blocks many carbon ions, therefore leading to a much longer coating process. Witness strips hanging off the outer diameter of the UCN guide showed coating step heights around $~\sim$75~nm. The coating on the inside of the tube is $\sim$20\% higher, which was experimentally determined by comparing steps heights between witness strips located along the inner diameter of tubes samples to those fixed to the outer diameter of the tube samples. Thus the coating on UCN guide should have a thickness around 90 nm which leads to a deposition rate in thickness per unit time of the UCN guide tube coating per length of $\sim12~nm$ $\cdot$h$^{-1}$m$^{-1}$. The flange coating was accomplished by spinning the flange above plasma plume at a rate of $\sim15$ rpm for 30 min, also without a plume collimator or TOF ion probe feedback. The plume geometry was such that the center of the flange was always intersecting the plume, but the edge of the flange would be rotated out of the plume. Thus we suspect a thickness gradient for the flange, which in general not a problem for UCN reflection provided the coating is thick enough to prevent tunneling. The most outer edge witness strip showed a thickness of 130 nm.

\begin{table}
\begin{centering}
\caption{Table detailing the coating parameters discussed in section \ref{sec:firstcoatings}.}
\label{tbl:samples}
    \begin{tabularx}{\linewidth}{|X|X|X|X|X|}

    \hline
    \textbf{Sample} & \textbf{Collimator} & \textbf{TOF Probe plume energy } &\textbf{Target bias} & \textbf{Coating adhered} \\
    \hline
     UCN Guide& none used & not used & -200 V & yes\\
    \hline
    Flange &  none used & not used & -200 V & yes\\
    \hline
     Tube sample &  oval & 100 eV & 0 V & no\\
     \hline
    \end{tabularx}
    \end{centering}
\end{table}

\subsection{Higher Density DLC coatings}
\begin{figure}[H]
    \centering
    \includegraphics[width=0.34\linewidth]{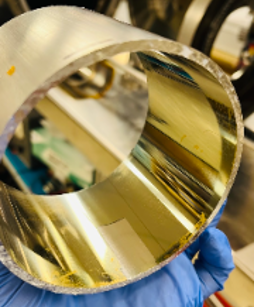}
    \caption{Photograph the DLC coated aluminum tube sample using TOF ion probe feedback and the plume collimator. The golden color is indicative of a more diamond-like coating. Profilometry preformed on its witness strips showed a thickness of 70 nm. After 24 hrs, the coating started to delaminate on spots of the tube sample. One can also see the silver rectangle where a Si witness strips was located.}
    \label{fig:betDLC}
\end{figure}

The density of the commissioning coatings via X-ray Reflectometry (XRR) on the witness strips, described in section \ref{sec:XRR}, showed low density more in line with graphite than diamond. It is then that we moved to implementing the TOF ion probe and plume collimator described in sections \ref{sec:TOF} and \ref{sec:collimator}. Figure \ref{fig:betDLC} shows a photograph of the tube sample with with these changes. It's golden color is indicative of a more diamond-like coating. Here we required the plume to have a TOF ion probe signal to correspond to $C^+$ ions with $\sim$100 eV energy. The target bias was 0 V, as to not influence the energy of the plume. Unfortunately, parts of the coating on the tube sample delaminated within 24 hrs, presumably from the higher internal stresses of the coating and weak adhesion to the aluminum substrate. However, some of the Si witness strips for these coatings did not delaminate and were investigated with XRR, which is discussed below, and stylus profilometry. The profilometer measurements showed a thickness of $\sim $75~nm which corresponds to a deposition rate in thickness per unit time of the coating per length of the cylindrical guide of 4 nm$\cdot$h$^{-1}$m$^{-1}.$ We note this is 1/3 deposition rate without the collimator. Table \ref{tbl:samples} lists the coating configurations used between the three samples.

\subsection{X-ray reflectometry Measurements}\label{sec:XRR}

\begin{figure}[H]
    \centering
    \includegraphics[width=.6\linewidth]{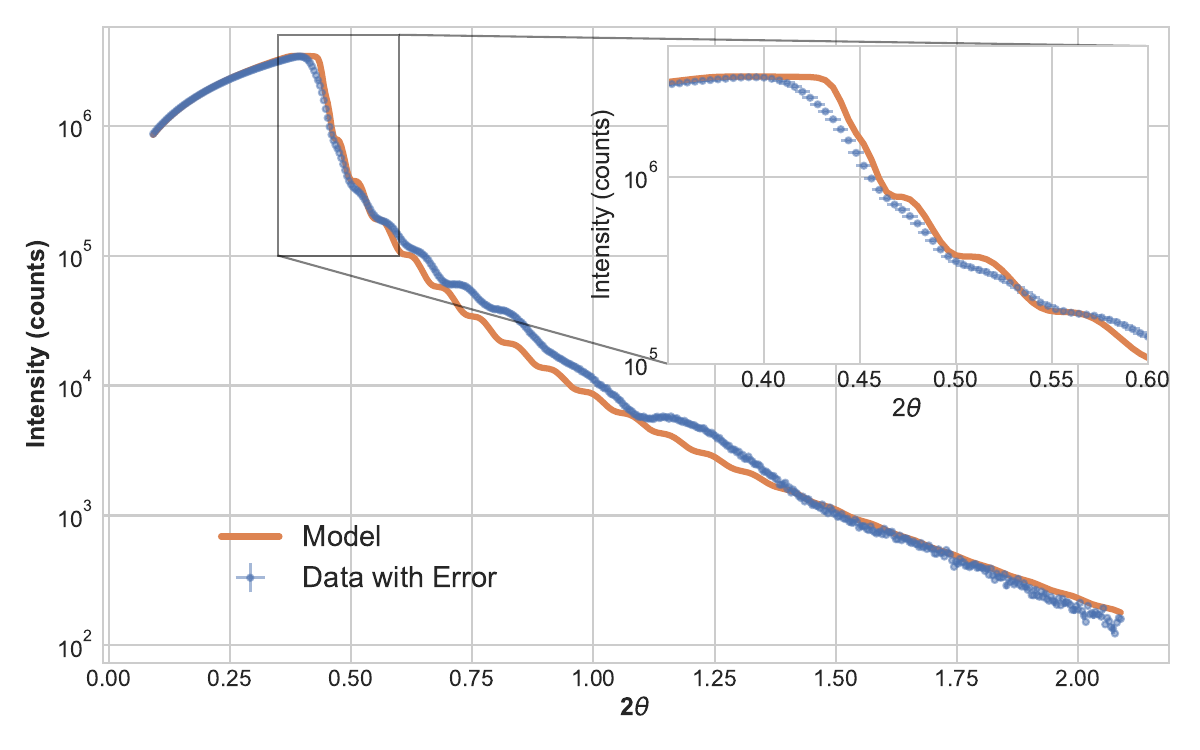}
    \includegraphics[width=.6\linewidth]{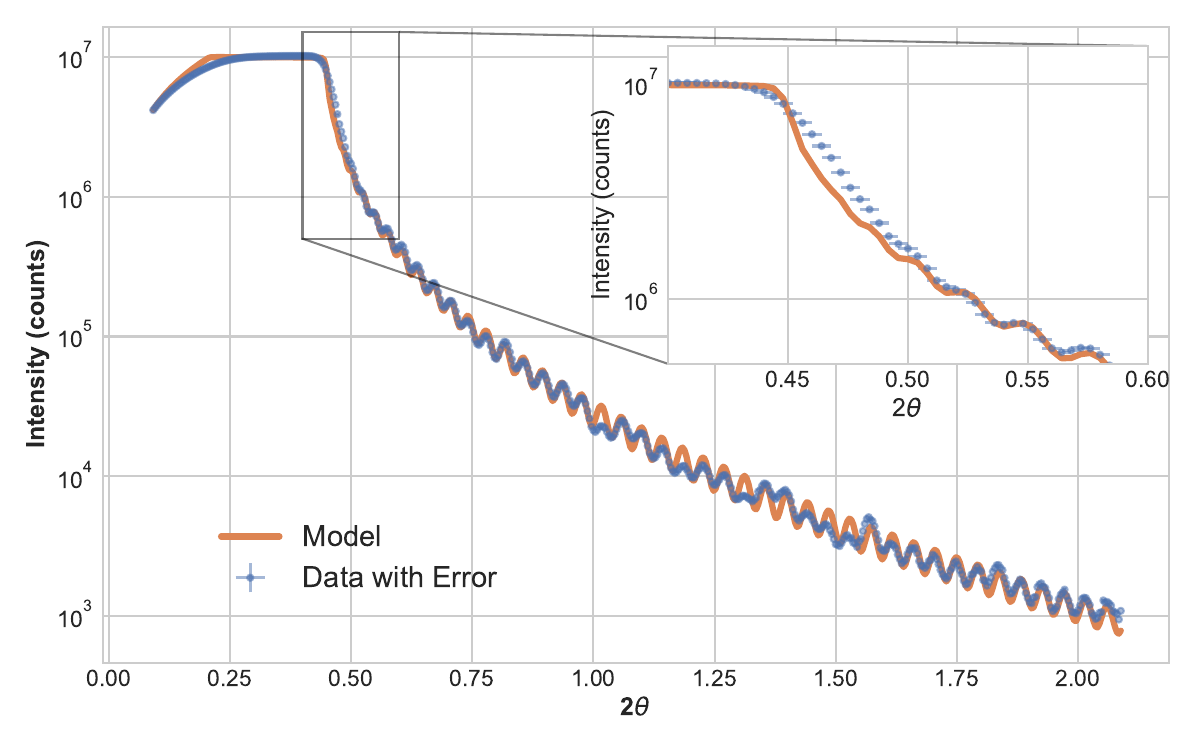}
     \includegraphics[width=.6\linewidth]{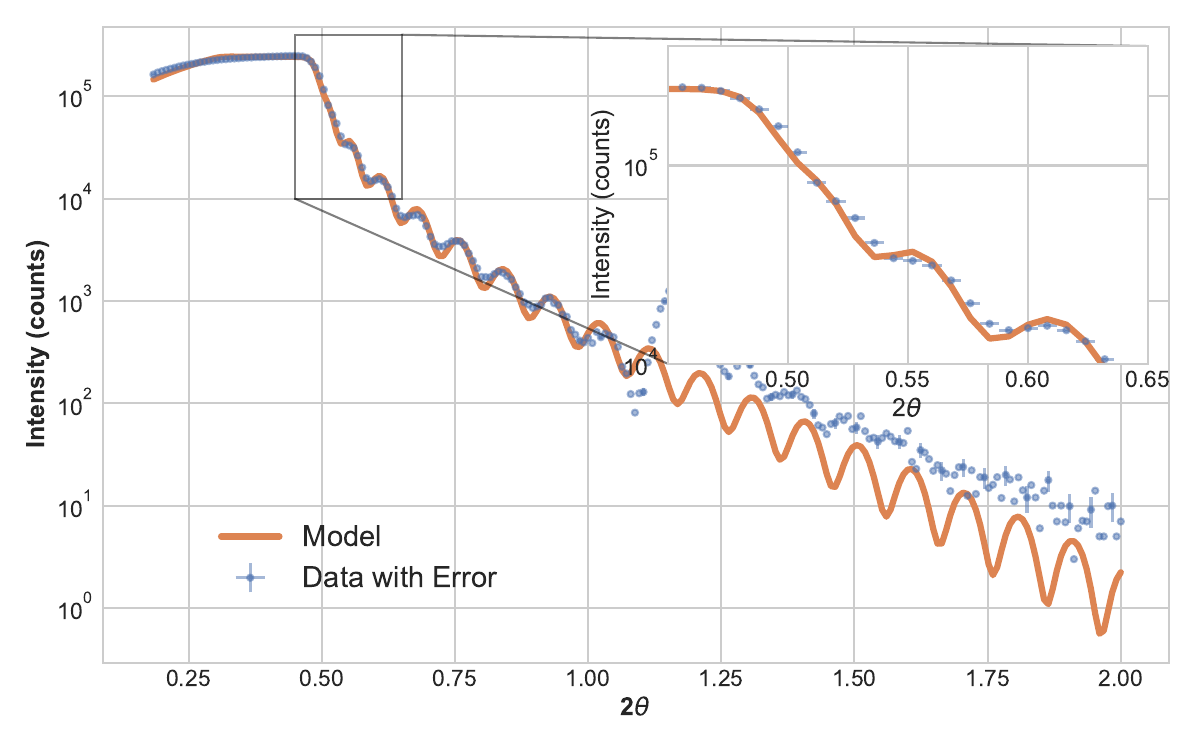}
    \caption{XRR data fits on silicon wafer witness strips via Rigaku SmartLab Studio II for three different coatings: (top) meter long UCN guide, flange (middle), and tube sample with the TOF ion probe feedback and plume collimator (bottom).  The data is in blue, while the model fits are in orange. The data was fit over the entire range shown, except for the bottom high density sample, where the data was only fit up to 2$\theta$=1.0. Insets, show the total reflection critical angle region.      }
    \label{fig:XRR}
\end{figure}

\begin{table}
\caption{XRR carbon fit parameters from witness strips for the three different coatings shown in \ref{fig:XRR}}
\label{tbl:XRR}
    \begin{tabularx}{\linewidth}{|c|c|c|c|c|}

    \hline
    \textbf{Sample} & \textbf{Carbon Density (g/cm$^3$}) & \textbf{Carbon thickness (nm) } &\textbf{Carbon Roughness (nm)} & \textbf{R factor $\%$}\\
    \hline
     UCN Guide& 2.28$\pm$.01 & 98.5 $\pm$.5 & .65 $\pm$ .01 & 3.4\\
    \hline
    Flange & 2.31$\pm$.01 & 191.8 $\pm$.1 & .42 $\pm$ .01 & 1.0\\
    \hline
     Tube sample & 2.82$\pm$.05 & 83.8 $\pm$.9 & 1.4 $\pm$ 1.6 & .5\\
     \hline
    \end{tabularx}
\end{table}

\subsection{Future Improvements}
The goal for the facility will be to increase the sp$^3$ bonding fraction and adhesion to the substrate of the coatings, and we are encouraged by the improvements in density seen with using TOF probe feedback and the plume collimator. Currently we are investigating interface coatings between the substrate and DLC, such as titanium or chromium, to promote adhesion. \\

It is also necessary to measure the imaginary component of the neutron-optical potential of our coatings. To measure this, a guide must be tested in a UCN storage or transmission experiment. This will occur after further optimization of the real component of the neutron-optical potential. After making a sufficiently optimized coating, the goal will shift to establishing reproducibility of coatings, and coating all guides for the TUCAN experiment. In addition, a vacuum deposition chamber dedicated to coated large flat substrates, up to 30" in diameter, will be commissioned in the near future. This system is intended to provide DLC coatings on the nEDM cell electrodes, deemed critical to the TUCAN experiment. 

\section{Conclusion}
A new facility for producing UCN-reflective coatings has been established and commissioned at University of Winnipeg. Using PLD, the facility applies DLC coatings to a variety of UCN hardware components, with a focus on 1 m long 95 mm inner diameter cylindrical UCN guides. The facility employs a 248 nm excimer laser, a large coating chamber, and a target rastering system to make the coatings.  \\

The first coatings of a 1 m aluminum guide and matching flange were completed, demonstrating the mechanical stability of the facility. Film density was measured to be (2.3)~g/cm$^3$ for the UCN guide and flange using XRR. These correspond to neutron-optical potentials of 200 neV. This confirms that some small amount of diamond bonding is present in the coatings, which were done without plume collimation or TOF ion probe plume feed back control. These films were measured to be 90-180 nm thick, and did not show any signs of delamination from the aluminum after one year. The implementation of a TOF ion probe to monitor the PLD plasma plume in concert with a plume collimator did produce a film with increased density (2.8~g/cm$^3$), corresponding to a 240~neV real optical potential, albeit with a significant drop in deposition rate and adhesion to the aluminum substrate. These results establish a baseline performance for the facility, and provide a foundation for future systematic studies of coating parameters to be completed to further increase coating diamond content and adhesion to the substrate. Optimized, high sp$^3$ fraction DLC coatings will directly benefit UCN transport and storage for the TUCAN experiment at TRIUMF, by delivering the maximum number of UCNs from their world-leading UCN source, to experiments. 

\section{Acknowledgements}
We would like to thank Dave Ostapchuk for his help settings things up at U.Winnipeg. We would like to thank Bruce Vogelar, Mark Pitt, Albert Young, Mark Makela, Brittney Vorndick, Grant Palmquist, Robert Pattie, Brian Dickerson and Xinjian Ding for their work at Virgina Tech. We also thank the UCNA collaboration for the support of the facility while at Virgina Tech. Vicky Jarvis at the MAX Diffraction facility at McMaster University provided the XRR analysis.\\

We gratefully acknowledge the support of the Canada Foundation for
Innovation; the Canada Research Chairs program; the Natural Sciences
and Engineering Research Council of Canada (NSERC) SAPPJ-2016-00024,
SAPPJ-2019-00031, SAPPJ-2023-00029, and SAPPJ-2024-00030; British Columbia Knowledge
Development Fund; Research Manitoba; JSPS KAKENHI (Grant
Nos. 18H05230, 19K23442, 20KK0069, 20K14487, and 22H01236, 25H00652); JSPS
Bilateral Program (Grant No. JSPSBP120239940); JST FOREST Program
(Grant No.  JPMJFR2237); International Joint Research Promotion
Program of Osaka University; RCNP COREnet; the Yamada Science Foundation; 
the Murata Science Foundation; the Grant for Overseas Research by the Division of Graduate Studies (DoGS) of Kyoto University; 
and the Universidad Nacional Aut\'onoma de M\'exico -
DGAPA program PASPA and grant PAPIIT AG102023.

\bibliographystyle{elsarticle-num}
\bibliography{UCN_references}

\end{document}